\documentstyle[aps,twocolumn]{revtex}

\begin{document}
\draft
\title{
Quantum state transfer between motion and light
}

\author{A.S. Parkins}
\address{
Department of Physics, University of Auckland, Auckland, New Zealand
}
\author{H.J. Kimble}
\address{
Norman Bridge Laboratory of Physics 12-33, California Institute of
Technology, Pasadena, CA 91125, U.S.A.
}

\maketitle

\begin{abstract}
We describe schemes for transferring quantum states between light
fields and the motion of a trapped atom. 
Coupling between the motion and the light is achieved via Raman
transitions driven by a laser field and the quantized field of
a high-finesse microscopic cavity mode. By cascading two such
systems and tailoring laser field pulses, we show that it is
possible to transfer an arbitrary motional state of one atom
to a second atom at a spatially distant site.
\end{abstract}


\narrowtext

\section{
Introduction
}

The quantized motional states of atoms or ions in confining
potentials offer interesting possibilities for a variety of
applications, such as the preparation and study of nonclassical
(i.e., manifestly quantum) states 
\cite{Meekhof96,Monroe96,Leibfried96,Cirac96,Deutsch98}, 
and the storage and manipulation
of quantum  information (e.g., ``qubits''), with particular
reference to quantum  logic operations and quantum computing
\cite{Cirac95a,Monroe95,King98,Turchette98a,Hughes96,Steane97,Wineland98}.
These possibilities stem from the relatively long coherence times 
that can be achieved  with motional states (due to the absence of 
strong damping mechanisms) and the precision with which 
transformations between motional states can be controlled using 
laser-light-induced transitions.

However, while motional states are well-suited to the storage and
manipulation of quantum states, for the communication of quantum
information from one physical location to another it is clear
that photons are the preferred carriers of the information.
For this reason, it is necessary to provide and  examine
configurations in which motional states can be efficiently and
reliably transferred to states of light, and vice-versa. Here
enters the field of cavity quantum electrodynamics (cavity QED);
in particular, configurations in which a single mode of the
electromagnetic field supported by an optical cavity is strongly
coupled to a transition in a single atom. It is possible, via
the internal atomic transition, to also couple the cavity field
to the external (quantized) motion of the trapped atom or ion
\cite{Zeng94,Buzek97,Harrison97},
and in this  work we will examine such a coupling that enables
the above-mentioned state transfer.

\section{
Model
}

Our model consists of a single two-level atom (or ion) confined
in a harmonic trap located inside an optical 
cavity. The atomic transition of frequency
$\omega_{\rm a}$ is coupled to a single mode of the cavity field of
frequency $\omega_{\rm c}$ and is also assumed to be driven by an
external (classical) laser field of frequency $\omega_{\rm L}$
-- the cavity and laser field frequencies  will be chosen so as to 
drive Raman transitions that couple  neighbouring vibrational levels
of the external motion. The physical setup and excitation scheme are
depicted in Fig.~1. The cavity is aligned along the $x$-axis, while
the laser field is incident from a direction in the $y$-$z$ plane
(i.e., perpendicular to the $x$-axis).

The Hamiltonian describing the internal and external atomic degrees
of freedom plus the atom-cavity and atom-laser couplings takes the  
form (in a frame rotating at the laser frequency)
\begin{eqnarray}
\hat{H}_0 = && 
\sum_{j=x,y,z} 
\hbar\nu_j (\hat{b}_j^\dagger\hat{b}_j+1/2) + \hbar\delta
\hat{a}^\dagger\hat{a} + \hbar\Delta\hat{\sigma}_+\hat{\sigma}_-
\nonumber
\\
&& \;\;\; +\, \hbar 
\left[ {\cal E}_{\rm L}(\hat{y},\hat{z},t) 
\hat{\sigma}_+ + 
{\cal E}_{\rm L}(\hat{y},\hat{z},t)^\ast
\hat{\sigma}_- \right] \nonumber
\\
&& \;\;\; +\, \hbar g_0 \sin (k\hat{x}) (\hat{a}^\dagger
\hat{\sigma}_- + \hat{\sigma}_+\hat{a} ) .
\end{eqnarray}
Here, $\{\nu_x,\nu_y,\nu_z\}$ are the harmonic oscillation frequencies
along the principal axes of the trap, 
$\hat{b}_j$ and $\hat{a}$ are annihilation operators for the
quantized atomic motion and cavity field, respectively, $\hat{\sigma}_-
=|g\rangle\langle e|$ is the atomic lowering operator, and
$\delta =\omega_{\rm c}-\omega_{\rm L}$ and 
$\Delta =\omega_{\rm a}-\omega_{\rm L}$. 
The quantity ${\cal E}_{\rm L}(\hat{y},\hat{z},t)$ is the (possibly
time-dependent) amplitude of the laser field; note again that we assume 
that this field has no spatial dependence along the $x$ direction.
Finally, the single-photon atom-cavity dipole coupling strength is
given by $g_0$, while the sine function describes the standing wave
structure of the cavity field (we assume that the centre of the trap
is located at a {\em node} of the cavity field), with $k=2\pi /\lambda$
the wavenumber of the field and $\hat{x}=[\hbar /(2m\nu_x)]^{1/2}
(\hat{b}_x+\hat{b}_x^\dagger )$. 

Allowing for cavity damping and atomic spontaneous emission, quantum
Langevin equations for the system operators can be derived
straightforwardly. For the moment, however, we will ignore the effects
of atomic spontaneous emission on the grounds that the detunings of 
the laser and cavity fields from the atomic transition frequency are 
very large, and hence that population of the excited atomic state 
$|e\rangle$ is negligible (we will return to the effects of 
spontaneous emission in the discussion at the end of the paper).
On this basis, we are also able to adiabatically eliminate
the internal atomic dynamics from the problem. 

We will also ignore any forms of motional decoherence or heating
associated with imperfections in the trap itself \cite{Wineland98} 
on the basis that such effects occur on a timescale slow compared 
with the operations we will be considering. Again, we will return to 
this point and discuss it more quantitatively at the end of the paper.

Finally, we assume that the size of the harmonic trap (in all 
directions) is small compared to the optical wavelength; 
under these conditions, we can make the approximation
$\sin (k\hat{x})\simeq \eta_x (\hat{b}_x+\hat{b}_x^\dagger )$, where
$\eta_x$ ($\ll 1$) is the Lamb-Dicke parameter. 
Given this assumption, it is also possible to design a configuration
for which we can neglect all position dependence in the laser field
\cite{laserantinode}; 
that is, we can assume a situation where
${\cal E}_{\rm L}(\hat{y},\hat{z},t)\simeq
{\cal E}_{\rm L}(t)e^{-i\phi_{\rm L}}$ [with ${\cal E}_{\rm L}(t)$
a real quantity].
Henceforth, the problem essentially becomes one-dimensional
and we can restrict our attention to just the $x$ direction.

To first order in $\eta_x$, equations of motion for the operators 
$\hat{a}$ and $\hat{b}_x$ then follow as
\begin{eqnarray}
\dot{\hat{a}} &=& -(\kappa +i\delta )\hat{a} 
+ i \frac{g_0\eta_x {\cal E}_{\rm L}
(t)e^{-i\phi_{\rm L}}}{\Delta} \, (\hat{b}_x+\hat{b}_x^\dagger ) 
\nonumber
\\
&& \;\;\;\;\;\; - \sqrt{2\kappa}\,
\hat{a}_{\rm in}(t) \; ,
\\
\dot{\hat{b}}_x &=& -i\nu_x\hat{b}_x 
+ i\frac{g_0\eta_x {\cal E}_{\rm L}(t)}{\Delta}
\, \left( \hat{a}^\dagger e^{-i\phi_{\rm L}} 
+ \hat{a} e^{i\phi_{\rm L}} \right) \; ,
\end{eqnarray}
where $\kappa$ is the decay rate of the cavity field and 
$\hat{a}_{\rm in}(t)$ is a quantum noise operator describing the 
input to the cavity field and satisfying the commutation relation 
$[\hat{a}_{\rm in}(t),\hat{a}_{\rm in}
^\dagger (t^\prime )] =\delta (t-t^\prime )$. 

We now make the transformation $\hat{b}_x=e^{-i\nu_x t}\tilde{b}_x$,
$\hat{a}=e^{-i\nu_x t}\tilde{a}$, and choose $\delta =\nu_x$ 
(i.e., tune to the ``first lower sideband''). 
Assuming that 
$\nu_x\gg\kappa$, $|(g_0\eta_x /\Delta ){\cal E}_{\rm L}(t)|$, and
$|\dot{{\cal E}}_{\rm L}(t)/{\cal E}_{\rm L}(t)|$, 
the oscillating terms in the resulting
equations may be dropped in a rotating-wave approximation to yield
\begin{eqnarray}
\dot{\tilde{a}} &=& -\kappa \tilde{a} 
+ i\frac{g_0\eta_x {\cal E}_{\rm L}(t)
e^{-i\phi_{\rm L}}}{\Delta} \, \tilde{b}_x 
- \sqrt{2\kappa}\, \tilde{a}_{\rm in}(t) 
\\
\dot{\tilde{b}}_x &=& i\frac{g_0\eta_x {\cal E}_{\rm L}(t)}{\Delta}
\tilde{a} e^{i\phi_{\rm L}} .
\end{eqnarray}
These equations simply describe a pair of coupled harmonic
oscillators, one of which is damped. In terms of a Hamiltonian, the
coupling is given by [defining 
$\Omega (t)=-g_0\eta_x {\cal E}_{\rm L}(t)/\Delta$ (real)]
\begin{equation}
\tilde{H}(t) = \hbar \Omega (t) \left( \tilde{a}^\dagger
\tilde{b}_x e^{-i\phi_{\rm L}} + \tilde{b}_x^\dagger \tilde{a} 
e^{i\phi_{\rm L}} \right) \; ,
\end{equation}
a result derived by Zeng and Lin \cite{Zeng94}.

\section{
Quantum state transfer
}

As pointed out by Zeng and Lin \cite{Zeng94}, 
when cavity losses can be neglected
the above coupling enables complete (pure or mixed) state
transfer between the atomic motion and the cavity light 
field. For example, given a finite laser pulse duration,
and assuming for simplicity that ${\cal E}_{\rm L}(t)$ is
a real and positive function of time, then solutions for
$\tilde{a}(t=+\infty )\equiv\tilde{a}(+\infty )$ and 
$\tilde{b}_x(t=+\infty )\equiv\tilde{b}_x(+\infty )$ can be
derived as
\begin{eqnarray}
\tilde{a}(+\infty ) &=& \tilde{a}(-\infty ) \cos\theta
- ie^{-i\phi_{\rm L}}\tilde{b}_x(-\infty ) \sin\theta 
\\
\tilde{b}_x(+\infty ) &=& - ie^{i\phi_{\rm L}}\tilde{a}(-\infty ) 
\sin\theta + \tilde{b}_x(-\infty ) \cos\theta \; ,
\end{eqnarray}
where $\theta =\int_{-\infty}^\infty \Omega (\tau )d\tau$. 
Choosing $\theta =(N+1/2)\pi$, with $N$ an integer, 
and $\phi_{\rm L} =\pi /2$ yields
\begin{equation}
\tilde{a}(+\infty ) = \mp \tilde{b}_x(-\infty ) \; , \;\;\;\;
\tilde{b}_x(+\infty ) = \pm \tilde{a}(-\infty ) \; ,
\end{equation}
from which it follows that, given an initial vacuum state of the
field, {\em any initial state of the motion can be transferred 
one-to-one to the state of the cavity field} and the motion is reduced 
to its ground state (for other examples of this kind of state exchange
between harmonic oscillator modes, see \cite{Steinbach97,Heinzen90}).

\subsection{
Underdamped regime
}

The result derived above demonstrating the possibility of 
complete state transfer between the quantized atomic motion and 
the cavity field mode obviously offers some very interesting 
further possibilities. If the damping of the cavity mode is 
sufficiently weak, then one can imagine a situation in which a 
suitable laser pulse is applied so as to transfer a motional state 
to the cavity mode (in a time short compared to $\kappa^{-1}$), 
after which the cavity field is allowed to decay. Making homodyne 
measurements on the output field from the cavity and using the 
well-established method of optical homodyne tomography 
\cite{Vogel89,Smithey93}, the density matrix of the cavity field 
mode, and hence of the initial motional state, could be 
reconstructed from many repeated cycles of preparation and 
measurement.

Given that the motion is always left in its ground state after the 
laser pulse, this can also be seen as a novel means of cooling, or 
``resetting,'' the atomic motion in a single operation
\cite{Heinzen90b}. Of course, in the regime of operation that we 
are assuming (the resolved sideband limit), conventional sideband 
cooling (using atomic spontaneous emission) would also be an 
efficient means of cooling the motion to the ground state.

\subsection{
Overdamped regime
}

The opposite limit, in which $\kappa$ is large compared to the 
magnitude of the effective coupling rate $\Omega (t)$
(but, of course, still small compared to the trap frequency 
$\nu_x$) is actually of more interest to us and indeed allows 
further simplification of the model. In particular, we can consider 
adiabatically eliminating the cavity mode from the dynamics, i.e., 
setting $\dot{\tilde{a}}=0$ and substituting 
\begin{equation} \label{adiab_elim}
\tilde{a} \simeq -\,\frac{\Omega (t)}{\kappa} \, \tilde{b}_x 
- \sqrt{2/\kappa}\, \tilde{a}_{\rm in}(t)
\end{equation}
into the equation for $\dot{\tilde{b}}_x$ to give
\begin{eqnarray}
\dot{\tilde{b}}_x &\simeq & - \frac{\left[ \Omega (t) \right]^2}{\kappa}
\tilde{b}_x + \Omega (t) \sqrt{2/\kappa}\, \tilde{a}_{\rm in}(t) 
\nonumber
\\
& \equiv & - \Gamma (t) \tilde{b}_x + \sqrt{2\Gamma (t)} \,
\tilde{a}_{\rm in}(t)\; ,
\label{eq:btilde}
\end{eqnarray}
where we have set $\phi_{\rm L}=\pi /2$ for simplicity.

This equation simply describes a quantum harmonic oscillator subject
to damping at the (possibly time-dependent) rate $\Gamma (t)$.
In the case of a vacuum cavity input field, it obviously models 
sideband cooling to the ground state due to a form of cavity-induced
spontaneous emission \cite{Cirac95b,Blatt99}.

\subsubsection{
Driven cavity: light-to-motion state transfer
}

However, one can also consider different kinds of inputs to the 
cavity field, i.e., choices of the input field operator 
$\tilde{a}_{\rm in}(t)$ that give rise to nontrivial input
field statistics. This is of interest because, given the simple
linear form of (\ref{eq:btilde}), it follows that the statistics of
the input field can be ``written onto'' the state of the oscillator.
In particular, assuming that $\Gamma (t)=\Gamma$, a constant, 
then in frequency space the solution to (\ref{eq:btilde}) is simply
\begin{equation}
\tilde{b}_x(\omega ) = \frac{\sqrt{2\Gamma}\,
\tilde{a}_{\rm in}(\omega )}{i\omega -\Gamma} \, .
\end{equation}

For example, the input field could be an 
{\em ideal quantum squeezed vacuum} as derived from the output of a 
degenerate parametric amplifier \cite{Kimble92,Turchette98b}. 
If the squeezing happens to be
broadband (with respect to the characteristic rates
associated with the system upon which it is incident), then the
appropriate input field correlation functions can be written in the
forms \cite{Gardiner91}
\begin{eqnarray}
\langle \tilde{a}_{\rm in}^\dagger 
(\omega )\tilde{a}_{\rm in}(\omega^\prime )
\rangle &=& N \, \delta (\omega -\omega^\prime ) 
\\
\langle \tilde{a}_{\rm in}(\omega )
\tilde{a}_{\rm in}(\omega^\prime )
\rangle &=& M \, \delta (\omega +\omega^\prime ) \, ,
\end{eqnarray}
with $M=|M|e^{i\theta}$ and $|M|^2=N(N+1)$. 
Given such an input, the system is equivalently
described by the master equation \cite{Gardiner91}
\begin{eqnarray} \label{eq:drhomdt}
\dot{\rho}_{\rm m} = && 
\Gamma (N+1)(2\tilde{b}_x\rho_{\rm m}\tilde{b}_x^\dagger -
\tilde{b}_x^\dagger\tilde{b}_x\rho_{\rm m} 
- \rho_{\rm m}\tilde{b}_x^\dagger\tilde{b}_x) \nonumber
\\
&& + \Gamma N(2\tilde{b}_x^\dagger\rho_{\rm m}\tilde{b}_x - 
\tilde{b}_x\tilde{b}_x^\dagger\rho_{\rm m} - 
\rho_{\rm m}\tilde{b}_x\tilde{b}_x^\dagger ) \nonumber
\\
&& - \Gamma M(2\tilde{b}_x\rho_{\rm m}\tilde{b}_x - 
\tilde{b}_x\tilde{b}_x\rho_{\rm m} -
\rho_{\rm m}\tilde{b}_x\tilde{b}_x) \nonumber
\\
&& - \Gamma M^\ast
(2\tilde{b}_x^\dagger\rho_{\rm m}\tilde{b}_x^\dagger -
\tilde{b}_x^\dagger\tilde{b}_x^\dagger\rho_{\rm m} - 
\rho_{\rm m}\tilde{b}_x^\dagger \tilde{b}_x^\dagger ) \, ,
\end{eqnarray}
where $\rho_{\rm m}$ is the density operator for the motion of the 
trapped atom. In steady state, the density operator is that of an 
ideal squeezed state, that is
\begin{equation}
\rho_{\rm m}^{\rm ss} = \hat{S}|0\rangle\langle 0|\hat{S}^+ \, ,
\end{equation}
where $\hat{S}$ is the squeezing operator \cite{Walls94}, i.e.,
$\hat{S}\tilde{b}\hat{S}^+=\mu\tilde{b}_x +\nu\tilde{b}_x^\dagger$
with $\mu =(N+1)^{1/2}$ and $\nu =N^{1/2}e^{i\theta}$.

There are of course other ways of preparing such nonclassical 
states of the motion which have indeed already been implemented 
experimentally \cite{Meekhof96,Monroe96,Leibfried96}. These 
preparations have typically employed pulsed {\em classical} light 
fields to facilitate the required motional state transformations.
The above scheme is novel in that it involves the direct 
transfer of statistics from a {\em nonclassical} continuous-wave 
light field to the motional state of the trapped atom.

\subsubsection{
Numerical calculations
}

To numerically model state transfer between light and motion in 
the overdamped regime described above, we consider, as above, the 
cavity mode to be resonantly driven by squeezed light from a 
degenerate parametric oscillator, as illustrated in Fig.~2. 
For our simulations, we include the dynamics of the parametric 
oscillator using the cascaded systems formalism developed in
\cite{Gardiner93,Carmichael93} (further applications of the 
formalism are given in \cite{Gardiner94}). In particular, we model 
our system with the master equation
\begin{eqnarray} \label{eq:sqme}
\dot{\rho} &=& - \frac{i}{\hbar} [\tilde{H}_{ab}(t)+\tilde{H}_c,
\rho ] \nonumber
\\
&& +\, \kappa_a \left( 2\tilde{a}\rho\tilde{a}^\dagger -
\tilde{a}^\dagger\tilde{a}\rho - \rho\tilde{a}^\dagger\tilde{a} 
\right)
\nonumber
\\
&& +\, \kappa_c \left( 2\tilde{c}\rho\tilde{c}^\dagger -
\tilde{c}^\dagger\tilde{c}\rho - \rho\tilde{c}^\dagger\tilde{c} 
\right)
\nonumber
\\
&& -\, 2\sqrt{\kappa_a\kappa_c} \, \left( [\tilde{a}^\dagger ,
\tilde{c}\rho ] + [\rho\tilde{c}^\dagger ,\tilde{a}] \right) \, .
\end{eqnarray}
Here, $\tilde{H}_{ab}(t)$ describes the effective coupling between 
the vibrational motion of the trapped atom and the cavity light 
field as derived above. However, in addition we retain the rotating, 
or nonsecular terms, i.e., we take
\begin{eqnarray}
\tilde{H}_{ab}(t) &=& \hbar \Omega \left( \tilde{a}^\dagger
\tilde{b}_x e^{-i\phi_{\rm L}} + \tilde{b}_x^\dagger \tilde{a} 
e^{i\phi_{\rm L}} \right) \nonumber
\\
&+& \, \hbar \Omega \left( \tilde{a}^\dagger
\tilde{b}_x^\dagger e^{-i\phi_{\rm L}+2i\nu_x t} 
+ \tilde{a} \tilde{b}_x e^{i\phi_{\rm L}-2i\nu_x t} \right) \, .
\end{eqnarray}
The coupling parameter $\Omega$ is assumed to be a constant. 
The Hamiltonian $\tilde{H}_c$ models the parametric oscillator
(driven below threshold), taking the form
\begin{equation}
\tilde{H}_c = \frac{1}{2} i\hbar \left[ \epsilon^\ast\tilde{c}^2
-\epsilon (\tilde{c}^\dagger )^2 \right] \, ,
\end{equation}
where $\tilde{c}$ is the annihilation operator for the cavity mode 
of the parametric oscillator and $\epsilon =|\epsilon |e^{i\theta}$ 
is the amplitude of the coherent field driving the oscillator. 
The linewidth of the (one-sided) parametric oscillator cavity mode 
is $\kappa_c$. Finally, the last term in (\ref{eq:sqme}) describes 
the (unidirectional) coupling of the incoming squeezed light to the 
cavity mode. This coupling is assumed to be ideal. 

By using a truncated state basis, the master equation 
(\ref{eq:sqme}) is numerically propagated until a steady state is 
achieved. In fact, due to the time-dependent terms in 
$\tilde{H}_{ab}(t)$, only a quasi-steady state can in principle be 
achieved, but for the parameters we consider only a very weak 
time-dependence (i.e., a weak modulation) occurs. The elements of 
the reduced density matrix of the motional mode in the
steady state ($\rho_{\rm m}^{\rm sim}$)
are shown in Fig.~3 for the choice of parameters 
$\nu_x/\kappa_a=10$, $\Omega /\kappa_a =0.1$, $\kappa_c/\kappa_a=1$ 
and $\epsilon /\kappa_c=0.3$ 
(corresponding to 71\% maximum squeezing in the input light field).
These parameters are in the regime for which the master equation
(\ref{eq:drhomdt}) should be valid and, indeed, we find the steady 
state of the motion to be approached on a timescale 
$\Gamma^{-1}=(\Omega^2/\kappa_a)^{-1}=100\kappa_a^{-1}$. 
Characteristic squeezed state features are evident in the figure 
(e.g., only even number states are populated) and the fidelity with 
which the predicted ideal squeezed state is achieved is computed to 
be 
$\langle 0|\hat{S}^+\rho_{\rm m}^{\rm sim}\hat{S}|0\rangle 
\simeq 0.99$
for the appropriate values of $N$ and $M$ in the theory.

\subsection{
Entanglement transfer from light fields to separated
trapped atoms
}

The scheme outlined above can be extended to the transfer of
quantum mechanical entanglement from light fields to motional 
states of two or more trapped atoms at physically separated sites.
Consider, for example, the pair of output fields from a 
{\em nondegenerate parametric amplifier} (the fields may be 
nondegenerate in polarization or in frequency) \cite{Kimble92}. 
At the output from the parametric amplifier, 
these fields could be separated in space and then made to impinge 
upon two cavities containing trapped atoms in the configuration
described above.
The quantum mechanical correlations that exist between the two 
light fields generated by parametric downconversion could thus be 
transferred (in steady state) to correlations between motional 
states of trapped atoms at two distinct sites.

\subsection{
Quantum teleportation of motional states
}

An exciting recent development in the field of quantum 
communication has been the experimental investigation of schemes
for the teleportation of quantum states
\cite{Boschi98,Bouwmeester97,Furusawa98}.
Of particular interest in the present context is the demonstration
by Furusawa {\em et al}. \cite{Furusawa98} of unconditional 
quantum teleportation of optical coherent states using 
squeezed-state fields and entanglement of the sort discussed above. 
By incorporating cavities containing trapped ions in the state 
transfer configuration of this work (in the bad cavity limit),
it should be possible to employ the scheme of \cite{Furusawa98}
for the teleportation of {\em motional} states.

In particular, a motional state could be ``mapped'' onto a light
field which enters the configuration of \cite{Furusawa98} at the
``sending'' station. This field is teleported to a ``receiving'' 
station where it is made to impinge upon a second trapped-ion plus
cavity system in the state transfer configuration. The teleported 
light field is thus mapped onto the motional mode of the second 
trapped ion and teleportation of the motional state is completed.
We will examine teleportation of motional states in more detail 
in a future work.

\subsection{
Motion-to-light state transfer: generation of nonclassical 
output light fields
}

Given the variety of, and efficiency with which, nonclassical 
motional states of single trapped atoms have been experimentally
realized \cite{Meekhof96,Monroe96,Leibfried96}, 
it is worth noting the potential of our scheme as a {\em source} 
of nonclassical output light fields. The output light field is
related to the input and internal cavity fields by 
\cite{Gardiner91,Walls94}
\begin{equation}
\tilde{a}_{\rm out}(t) = \tilde{a}_{\rm in}(t) + \sqrt{2\kappa}\,
\tilde{a}(t) \, ,
\end{equation}
which in the overdamped limit becomes [using (\ref{adiab_elim})]
\begin{equation}
\tilde{a}_{\rm out}(t) \simeq  - \tilde{a}_{\rm in}(t) 
- \sqrt{2\Gamma (t)}\, \tilde{b}_x(t) \, ,
\end{equation}
where we assume that $\Omega (t)\geq 0$. Hence, given a vacuum
field input to the cavity, the output field is determined by the
motional state of the trapped atom. Further, 
depending on the nature of the motional state preparation, the 
output may be pulsed {\em or} continuous; for a continuous output
one would have $\Gamma (t)=\Gamma$, a constant, and the motional
state preparation scheme would have to operate in a continuous
manner also. As an example, consider squeezed motional states,
which may be generated by applying an electric field gradient
with a frequency $2\nu_x$ to the ion \cite{Heinzen90}, or by
irradiating the ion with two laser beams differing in frequency by
$2\nu_x$ \cite{Meekhof96}.

Note that in \cite{Meekhof96} squeezed states of the motion were 
produced exhibiting a reduction in the variance of the squeezed 
quadrature by a factor of 40. 
Such quadrature noise reduction has yet to be
approached via traditional optical means, suggesting that the
present state transfer configuration is worthy of further 
investigation \cite{LDlimit}.

\section{
Transfer of a motional state between separated trapped
atoms
}

Recently, Cirac {\em et al.} \cite{Cirac97} (see also
\cite{vanEnk97,Pellizzari97,vanEnk98}) demonstrated how 
quantum transmission of a qubit between two nodes of a 
quantum network can be implemented in a physical system
using light as the carrier of the quantum information. 
In particular, they showed how the transformation
\begin{equation} \label{qubittrans}
(c_0|0\rangle_1 + c_1|1\rangle_1) \otimes |0\rangle_2 
\rightarrow |0\rangle_1 \otimes 
(c_0|0\rangle_2 + c_1|1\rangle_2) 
\end{equation}
can be achieved where $|0\rangle_1$ and $|1\rangle_1$ are
internal states of an atom at node 1 and 
$|0\rangle_2$ and $|1\rangle_2$ are the corresponding states of
a second atom at (the spatially separated) node 2. 
At each node, the atom is located within a cavity supporting a
single mode of the electromagnetic field, with which it is made to 
undergo a controlled time-dependent interaction via a laser-assisted 
Raman process. With suitably chosen laser pulses at the two
nodes, the transmission described by (\ref{qubittrans}) can be
faithfully reproduced, facilitated by the transfer of a {\em photon} 
wave packet between the two nodes.

In the same spirit, we consider here the transmission of 
{\em arbitrary motional states} of trapped atoms between two distinct 
sites, facilitated once again by cavity light fields and photon 
wave packets. We consider two separated atom-cavity arrangements, 
each in the configuration described in Section II, so that the 
combined system Hamiltonian can be written as
\begin{eqnarray}
\tilde{H} &=& \hbar \Omega_1 (t) \left( \tilde{a}_1^\dagger
\tilde{b}_1 e^{-i\phi_1} + \tilde{b}_1^\dagger \tilde{a}_1
e^{i\phi_1} \right) \nonumber
\\
&& +
\hbar \Omega_2 (t) \left( \tilde{a}_2^\dagger
\tilde{b}_2 e^{-i\phi_2} + \tilde{b}_2^\dagger \tilde{a}_2
e^{i\phi_2} \right) \, ,
\end{eqnarray}
where the subscripts $\{ 1,2\}$ denote the site of each atom or
cavity mode and we now omit the subscript $x$ for simplicity.

\subsection{
Cascaded systems model
}

The two cavity modes are each damped at rate $\kappa$, but coupling 
between their external fields is assumed to be unidirectional. In
particular, the output from cavity 1 is incident upon (i.e., provides
the input field to) cavity 2, but not {\em vice versa}. 
Such a situation is depicted in Fig.~4 and 
is again modeled by the cascaded systems formalism 
introduced earlier \cite{Gardiner93,Carmichael93}. 
In this formalism, the master equation for our system is derived
in the form
\begin{equation}
\dot{\rho} = \left( {\cal L}_0 + {\cal L}_{\rm c} \right) \rho \, ,
\end{equation}
where
${\cal L}_0\rho = -(i/\hbar )[\tilde{H},\rho ]$ 
and
\begin{eqnarray} \label{Lc}
{\cal L}_{\rm c}\rho &=& 
\kappa \left( 2\tilde{a}_1\rho\tilde{a}_1^\dagger
- \tilde{a}_1^\dagger\tilde{a}_1\rho - \rho\tilde{a}_1^\dagger
\tilde{a}_1 \right) \nonumber
\\
&& + \kappa \left( 2\tilde{a}_2\rho\tilde{a}_2^\dagger
- \tilde{a}_2^\dagger\tilde{a}_2\rho - \rho\tilde{a}_2^\dagger
\tilde{a}_2 \right) \nonumber
\\
&& - 2\kappa \left( [\tilde{a}_2^\dagger ,\tilde{a}_1\rho ]
+ [\rho\tilde{a}_1^\dagger ,\tilde{a}_2 ] \right) ,
\end{eqnarray}
where a vacuum input to the first cavity has been assumed.
The last term in (\ref{Lc}) provides the desired unidirectional 
coupling in the theory. 

To simplify the model, we assume, as before, that the decay rate 
$\kappa$ is large compared to other rates in the system (apart 
from the trap frequency $\nu$) and that the cavity fields can be 
adiabatically eliminated from the system dynamics. In the approach 
we are following in this section, this leads to a reduced master 
equation for the density operator of the motion of the two atoms,
given formally by \cite{Gardiner91}
\begin{equation}
\dot{\rho}_{\rm m} = {\rm Tr}_{\rm c} \left\{ {\cal L}_0
\int_0^\infty d\tau\; e^{{\cal L}_{\rm c}\tau} {\cal L}_0 
\rho_{\rm c}^{\rm ss} \,\right\} \rho_{\rm m} ,
\end{equation}
where $\rho_{\rm c}^{\rm ss}$ is the steady state density matrix 
for the two cavity modes. To evaluate 
this expression explicitly requires steady state correlation
functions for the operators $\tilde{a}_1$ and $\tilde{a}_2$, which
can be derived from (\ref{Lc}) using the quantum regression theorem.
As shown in the appendix, 
the only nonzero correlation functions are
\begin{equation} \label{corr1}
\langle \tilde{a}_1(\tau )\tilde{a}_1^\dagger (0)\rangle_{\rm ss} =
\langle \tilde{a}_2(\tau )\tilde{a}_2^\dagger (0)\rangle_{\rm ss} =
e^{-\kappa\tau} 
\end{equation}
and
\begin{equation} \label{corr2}
\langle \tilde{a}_1(0)\tilde{a}_2^\dagger (\tau )\rangle_{\rm ss} =
\langle \tilde{a}_2(\tau )\tilde{a}_1^\dagger (0)\rangle_{\rm ss} =
-2\kappa\tau e^{-\kappa\tau} \; .
\end{equation}
Using these expressions gives
\begin{eqnarray}
\dot{\rho}_{\rm m} &=& \Gamma_1(t)
\left( 2\tilde{b}_1\rho_{\rm m}\tilde{b}_1^\dagger
- \tilde{b}_1^\dagger\tilde{b}_1\rho_{\rm m} 
- \rho_{\rm m}\tilde{b}_1^\dagger\tilde{b}_1 \right) \nonumber
\\
&& + \Gamma_2(t) \left( 2\tilde{b}_2\rho_{\rm m}\tilde{b}_2^\dagger
- \tilde{b}_2^\dagger\tilde{b}_2\rho_{\rm m} 
- \rho_{\rm m}\tilde{b}_2^\dagger\tilde{b}_2 \right) \nonumber
\\
&& + 2\sqrt{\Gamma_1(t)\Gamma_2(t)} 
\left\{ [\tilde{b}_2^\dagger ,\tilde{b}_1\rho_{\rm m} ]
e^{-i(\phi_1-\phi_2)} \right. \nonumber
\\
&& \;\;\;\;\;\;\;\;\;\; \left.
+ [\rho_{\rm m}\tilde{b}_1^\dagger ,\tilde{b}_2 ] 
e^{i(\phi_1-\phi_2)} \right\} \, .
\end{eqnarray}
This master equation once again describes a cascaded system, only
now the ``coupling'' appears directly between the motional modes
of the two atoms.

\subsection{
Quantum trajectories and ideal state transmission
}

To demonstrate the transmission properties of the coupled system, 
we again follow Cirac {\em et al.} and employ the technique of
quantum trajectories \cite{Carmichael93,Zoller95}. 
This technique simulates a given master equation by propagating a
system wave function $|\psi (t)\rangle$ subject to a non-Hermitian 
effective Hamiltonian. This propagation is interrupted at random 
times $\{ t_r\}$ by wave function collapses, or quantum jumps, 
$|\psi (t_r+dt)\rangle =\tilde{C}|\psi (t)\rangle$, which 
can be interpreted as, in our particular instance, the emission 
and destructive detection of photons from the cavity fields. 

For the master equation derived above, the effective Hamiltonian
takes the form (choosing the laser phases such that $\phi_1=\phi_2$)
\begin{eqnarray} \label{Heff}
\tilde{H}_{\rm eff}(t) &=& -i\Gamma_1(t)\tilde{b}_1^\dagger 
\tilde{b}_1 -i\Gamma_2(t)\tilde{b}_2^\dagger \tilde{b}_2  
\nonumber
\\
&& \;\;\; + 2i\sqrt{\Gamma_1(t)\Gamma_2(t)} \,
\tilde{b}_2^\dagger \tilde{b}_1 \, ,
\end{eqnarray}
while the collapse operator is given by
\begin{equation}
\tilde{C} = \sqrt{\Gamma_1(t)}\, \tilde{b}_1
- \sqrt{\Gamma_2(t)}\, \tilde{b}_2 \, .
\end{equation}

The basic idea is to design laser pulse profiles [manifest through
$\Gamma_i(t)$] at the two sites such that the ideal quantum
transmission
\begin{equation} \label{perftrans}
\sum_{n=0}^\infty c_n|n\rangle_1 \otimes |0\rangle_2
\rightarrow |0\rangle_1 \otimes \sum_{n=0}^\infty c_n|n\rangle_2 ,
\end{equation}
can be achieved. Here $|n\rangle_i$ denotes the $n$-th Fock state 
of the motion of atom $i$. A necessary condition for successful 
transmission is that a quantum jump never occurs, i.e., 
$\tilde{C}|\psi (t)\rangle =0$ for all $t$, in which case the 
effective Hamiltonian becomes a Hermitian operator. This can be 
interpreted in terms of transfer via a {\em dark} state of the 
cascaded system.

We expand the state of the system as
\begin{equation} \label{psi_t}
|\psi (t)\rangle = \sum_{n=0}^\infty c_n \sum_{m=0}^n
\alpha_m^{(n)}(t) |n-m\rangle_1 \otimes |m\rangle_2 ,
\end{equation}
with initial condition
\begin{equation}
\alpha_0^{(n)}(-\infty ) = 1 \, , \;\;\;\;
\alpha_{m\neq 0}^{(n)}(-\infty ) = 0 ,
\end{equation}
and normalization
\begin{equation}
\sum_{m=0}^n \left| \alpha_m^{(n)}(t) \right|^2 = 1 .
\end{equation}
For ideal quantum transmission, one requires that
\begin{equation}
\alpha_n^{(n)}(+\infty ) = \alpha_0^{(n)}(-\infty ) = 1 .
\end{equation}

Equations of motion for the $\{\alpha_m^{(n)}(t)\}$ are derived
using (\ref{Heff}) and (\ref{psi_t}); the equations 
for the $m=0$ components take the simple closed form
\begin{equation}
\dot{\alpha}_0^{(n)}(t) = -n\Gamma_1(t)\alpha_0^{(n)}(t) ,
\end{equation}
with solutions
\begin{equation}
\alpha_0^{(n)}(t) = \exp \left\{ 
-n\int_{-\infty}^t \Gamma_1(t^\prime ) dt^\prime \right\} \; .
\end{equation}
Now, applying the dark state condition 
\begin{equation}
\left\{ \sqrt{\Gamma_1(t)}\, \tilde{b}_1
- \sqrt{\Gamma_2(t)}\, \tilde{b}_2 \right\}
|\psi (t)\rangle = 0
\end{equation}
yields the sequence of straightforward algebraic equations
\begin{eqnarray}
\sqrt{n\Gamma_1(t)}\,\alpha_0^{(n)}(t) - \sqrt{\Gamma_2(t)}\,
\alpha_1^{(n)}(t) &=& 0 \, , \nonumber
\\
\sqrt{(n-1)\Gamma_1(t)}\,\alpha_1^{(n)}(t) -\sqrt{2\Gamma_2(t)}\, 
\alpha_2^{(n)}(t) &=& 0 \, , \nonumber
\\
\sqrt{(n-2)\Gamma_1(t)}\,\alpha_2^{(n)}(t) -\sqrt{3\Gamma_2(t)}\, 
\alpha_3^{(n)}(t) &=& 0 \, , \;\;\; \ldots \; ,
\end{eqnarray}
from which a solution for $\alpha_m^{(n)}(t)$ in terms of
$\alpha_0^{(n)}(t)$ follows as
\begin{equation}
\alpha_m^{(n)}(t) = \left[ \frac{\Gamma_1(t)}{\Gamma_2(t)} 
\right]^{m/2} \sqrt{\frac{n!}{m!(n-m)!}} \, \alpha_0^{(n)}(t) .
\end{equation}
Applying the normalization condition gives
\begin{equation}
\sum_{m=0}^n \left[ \alpha_m^{(n)}(t) \right]^2 =
\left[ 1 + \frac{\Gamma_1(t)}{\Gamma_2(t)} \right]^n 
\left[ \alpha_0^{(n)}(t) \right]^2 = 1
\end{equation}
and thus
\begin{equation} \label{Gam12}
\frac{\Gamma_1(t)}{\Gamma_2(t)} = 
\exp \left\{ 2\int_{-\infty}^t \Gamma_1(t^\prime )
dt^\prime \right\} - 1 ,
\end{equation}
which demonstrates that it is possible to choose
pulse shapes of the laser fields in such a way that the perfect
transmission (\ref{perftrans}) can be achieved.

We have not explored in detail pulse shapes satisfying
(\ref{Gam12}). We do find, however, that (\ref{Gam12}) 
admits the following simple {\em analytical} solutions,
\begin{equation}
\Gamma_1(t) = \Gamma \, 
\frac{e^{\Gamma t}}{e^{\Gamma t}+e^{-\Gamma t}} 
\, , \;\;\;\;
\Gamma_2(t) = \Gamma_1(-t) \, ,
\end{equation}
with the limits $\Gamma_1(t)\rightarrow 0$ ($\Gamma$) as 
$t\rightarrow -\infty\; (+\infty )$,
and {\em vice-versa} for $\Gamma_2(t)$. We use these forms
in the numerical calculations that follow.

\subsection{
Numerical calculations
}

For the purpose of numerical calculations, we retain the dynamics
of the cavity field modes and solve the cascaded systems master 
equation
\begin{eqnarray}
\dot{\rho} &=& - \frac{i}{\hbar} [\tilde{H}_1(t)+\tilde{H}_2(t),
\rho ] \nonumber
\\
&& +\, \kappa \left( 2\tilde{a}_1\rho\tilde{a}_1^\dagger -
\tilde{a}_1^\dagger\tilde{a}_1\rho 
- \rho\tilde{a}_1^\dagger\tilde{a}_1 \right)
\nonumber
\\
&& +\, \kappa \left( 2\tilde{a}_2\rho\tilde{a}_2^\dagger -
\tilde{a}_2^\dagger\tilde{a}_2\rho 
- \rho\tilde{a}_2^\dagger\tilde{a}_2 \right)
\nonumber
\\
&& -\, 2\kappa \, \left( [\tilde{a}_2^\dagger ,
\tilde{a}_1\rho ] + [\rho\tilde{a}_1^\dagger ,\tilde{a}_2] \right)
\end{eqnarray}
with
\begin{eqnarray}
\tilde{H}_k(t) &=& \hbar \Omega_k(t) \left( \tilde{a}_k^\dagger
\tilde{b}_k e^{-i\phi_k} + \tilde{b}_k^\dagger \tilde{a}_k 
e^{i\phi_k} \right) \nonumber
\\
&+& \, \hbar \Omega_k(t) \left( \tilde{a}_k^\dagger
\tilde{b}_k^\dagger e^{-i\phi_k+2i\nu t} 
+ \tilde{a}_k \tilde{b}_k e^{i\phi_k-2i\nu t} \right) \, ,
\end{eqnarray}
where $k=1,2$.
Once again, we include the rotating terms 
$\tilde{a}_k\tilde{b}_k$ and 
$\tilde{a}_k^\dagger\tilde{b}_k^\dagger$,
while the forms of the 
time-dependent effective coupling parameters
$\Omega_1(t)$ and $\Omega_2(t)$ are chosen in accordance with the
work of the previous section, i.e.,
\begin{equation} \label{Om_overdamped}
\Omega_1(t) = \Omega \, 
\sqrt{\frac{e^{\Gamma t}}{e^{\Gamma t}+e^{-\Gamma t}}}
= \Omega_2(-t) \, ,
\end{equation}
where $\Gamma = \Omega^2/\kappa$ (and we chosen 
$\phi_1=\phi_2=0$).

As the state to be transferred, we choose, arbitrarily 
(in practice we are somewhat limited by the size
of the basis set we can use comfortably in our simulations), 
\begin{equation} \label{psi}
|\psi\rangle = \frac{1}{2} \left( |0\rangle +
e^{i\pi /3}|1\rangle + e^{i2\pi /3}|2\rangle + e^{i\pi}|3\rangle
\right) \, ,
\end{equation}
so that the initial state of the total system is
$|0\rangle_{a1} \otimes |\psi\rangle_{b1} \otimes 
|0\rangle_{a2} \otimes |0\rangle_{b2}$, while the target state is
\begin{equation}
|\psi_{\rm target}\rangle = |0\rangle_{a1} \otimes |0\rangle_{b1}
\otimes |0\rangle_{a2} \otimes |\psi\rangle_{b2} \, .
\end{equation}
The transmission fidelity, which we define by
\begin{equation}
F(t) = 
\langle\psi_{\rm target}|\rho (t)|\psi_{\rm target}\rangle \, ,
\end{equation}
is plotted in Fig.~5 for $\Omega /\kappa =0.141$ (corresponding to
$\Gamma /\kappa =0.02$) and three different values of the trap 
frequency $\nu$. 
Note that the initial value of the fidelity is finite due to the
contribution from $|0\rangle$ in the state $|\psi\rangle$.
For $\nu /\kappa =20$ ($10$) the state is transmitted with a 
fidelity of 0.995 (0.980). As $\nu /\kappa$ is lowered the fidelity
is reduced as the rotating terms in the effective Hamiltonians 
$\tilde{H}_k(t)$ begin to contribute more strongly to the dynamics.
Nevertheless, a fidelity of 0.925 is found even for 
$\nu /\kappa =5$.

In Fig.~6 we consider a single value of the trap frequency, 
$\nu /\kappa =20$, but now vary the coupling parameter $\Omega$,
which varies the effective rate $\Gamma$ of the state transfer 
operation. We also choose larger values of $\Omega$ compared
to $\kappa$ to assess how well the chosen pulse shapes drive the
state transfer as the adiabatic approximation ($\kappa\gg\Omega$) 
ceases to be valid. As one can see, the fidelity of the 
transmission remains high even with $\Omega /\kappa =0.4$ (0.984)
and $0.5$ (0.970), but at $\Omega /\kappa =0.7$ deteriorates to
0.905. Notably, the timescales for the transfer are significantly
faster than in the previous figure, which is possibly 
advantageous from an experimental point of view.

\subsection{
Underdamped regime
}

The overdamped regime considered above probably corresponds to
the most likely experimental scenario. However, a situation where
$\Omega >\kappa$ is still possible and could offer some 
further advantages in terms of transfer rates, which, 
in such a regime, would be of the order of $\kappa$.
Of course, if $\Omega >\kappa$ then one must include the cavity
modes and their dynamics in the analysis, which thereby becomes
somewhat more complicated.

Nevertheless, we find that it is still possible to derive pulse 
shapes that allow high-fidelity transmission of arbitrary quantum 
states. In particular, 
following the kind of approach used in \cite{Cirac97}, whereby
one assumes that $\Omega_1(t)=\Omega$ (a constant) for $t\geq 0$
and that $\Omega_2(t)=\Omega_1(-t)$ (i.e., the 
{\em symmetric pulse condition}), we are able to arrive at
the form
\begin{equation} \label{eq:Om_underdamped}
\Omega_1(-t) = -\,\frac{\Omega f(t)+2\kappa h(t)}
{\sqrt{1-f(t)^2-2h(t)^2}} \;\;\;\; (t\geq 0),
\end{equation}
where
\begin{eqnarray}
f(t) &=& \frac{1}{p} \left( \lambda_+ e^{\lambda_-t} - \lambda_-
e^{\lambda_+t} \right) f(0) \nonumber
\\
&& \;\;\;\;\; + \frac{\Omega}{p} \left( 
e^{\lambda_+t} - e^{\lambda_-t} \right) h(0) \, ,
\\
h(t) &=& \frac{1}{p} \left( \lambda_+ e^{\lambda_+t} - \lambda_-
e^{\lambda_-t} \right) h(0) \nonumber
\\
&& \;\;\;\;\; - \frac{\Omega}{p} \left( 
e^{\lambda_+t} - e^{\lambda_-t} \right) f(0) \, ,
\end{eqnarray}
with
$\lambda_\pm = -(1/2)(\kappa \pm p)$, 
$p = (\kappa^2-4\Omega^2)^{1/2}$,
and
\begin{equation}
f(0) = \sqrt{\frac{\kappa^2/2}{\kappa^2+\Omega^2}} \, , \;\;\;\;
h(0) = -\sqrt{\frac{\Omega^2/2}{\kappa^2+\Omega^2}} \, .
\end{equation}
That such forms for the pulse shapes can successfully facilitate 
state transmission between the two atoms is illustrated by the
dotted line in Fig.~6
for the same state $|\psi\rangle$ considered above and with
$\Omega /\kappa =0.7$. The state is transmitted with a fidelity
of 0.993, clearly improving on the result from the approach
in the overdamped regime and also doing so on a faster timescale.

\section{
Discussion
}

In this paper we have described schemes for the transfer of quite
general quantum states between light fields and atomic motion and
between the motion of trapped atoms at separate sites. We want now 
to consider in more detail some of the basic assumptions involved
in our model and to examine possible experimental situations. 

Clearly, a very important assumption is that the effects of 
atomic spontaneous emission can be neglected. 
In a master equation approach, atomic spontaneous emission with the
effects of recoil taken into account is modelled by a term of the
form (considering motion only along the $x$ axis)
\cite{Cirac92}
\begin{equation}
\{\dot{\rho}\}_{\rm spon} = \frac{\gamma}{2} 
\left( 2\hat{\sigma}_-\tilde{\rho}\hat{\sigma}_+ - 
\hat{\sigma}_+\hat{\sigma}_-\rho - 
\rho\hat{\sigma}_+\hat{\sigma}_- \right) \, ,
\end{equation}
where
\begin{eqnarray}
\tilde{\rho} &=& \frac{1}{2} \int_{-1}^{+1} du \; W(u)
e^{iku\hat{x}} \rho e^{-iku\hat{x}} \nonumber
\\
&=& \frac{1}{2} \int_{-1}^{+1} du \; W(u)
e^{i\eta_x (\hat{b}_x+\hat{b}_x^\dagger )} \rho 
e^{-i\eta_x (\hat{b}_x+\hat{b}_x^\dagger )} \, .
\end{eqnarray}
Here, $\gamma$ is the spontaneous emission rate and 
$W(u)=(3/4)(1+u^2)$ describes the angular distribution of
spontaneous emission for an atomic dipole transition.
Incorporating this into our analysis and adiabatically
eliminating the atomic and cavity degrees of freedom as
before, one finds that the leading order (in $\eta_x$) 
contribution to the motional dynamics contributed by 
atomic recoil due to spontaneous emission takes the form
\begin{eqnarray}
\eta_x^2 \, \frac{\gamma}{10} \frac{{\cal E}_{\rm L}^2}{\Delta^2}
&& \left[ 
2(\hat{b}_x+\hat{b}_x^\dagger )\rho (\hat{b}_x+\hat{b}_x^\dagger )
\right. \nonumber
\\
&& \;\;\; - \left. (\hat{b}_x+\hat{b}_x^\dagger )^2\rho - 
\rho (\hat{b}_x+\hat{b}_x^\dagger )^2 \right] \, .
\end{eqnarray}
[Note that we also assume that 
${\cal E}_{\rm L}\gg\eta_x g_0[\langle\hat{a}^\dagger\hat{a}
\rangle ]^{1/2}$.]

From inspection of the plots of transmission fidelity versus
time (for the {\em overdamped} regime), we can estimate the 
timescale for state transfer as $\sim 4/\Gamma$. 
Hence, in order to be able to neglect the effects of spontaneous
emission on the transfer process, we require that  
\begin{equation} \label{SpEcondition}
\frac{\Gamma}{4} = 
\frac{\eta_x^2g_0^2{\cal E}_{\rm L}^2}{4\kappa\Delta^2} \gg
\eta_x^2 \, \frac{\gamma}{10} \frac{{\cal E}_{\rm L}^2}{\Delta^2}
\;\;\;\; {\rm or} \;\;\;\;
\frac{5g_0^2}{2\kappa\gamma} \gg 1 \, .
\end{equation}
This, not surprisingly, corresponds to the regime of 
strong-coupling in cavity QED \cite{Kimble94}.

The dipole coupling strength is given by
$g_0 = [3c\lambda^2\gamma /(8\pi V_{\rm m})]^{1/2}$,
where $\lambda$ is the wavelength of the atomic transition
and $V_{\rm m}=(\pi /4)w_0^2l$ is the cavity mode volume,
with $w_0$ the cavity mode waist and $l$ the mirror separation. 
The cavity field decay rate can be expressed in terms of the mirror
separation and the cavity finesse ${\cal F}$ as
$\kappa = \pi c/(2{\cal F}l)$. 
Using these expressions for $g_0$ and $\kappa$, the condition in
(\ref{SpEcondition}) can be rewritten in the form
\begin{equation}
\frac{15}{2} \, \frac{\lambda^2}{\pi^3} \, \frac{{\cal F}}{w_0^2} 
\gg 1  .
\end{equation}
So, of course, one would like to have small cavity modes and
high finesse mirrors. 

The trap itself must also meet rather stringent requirements;
in particular, the Lamb-Dicke parameter must 
satisfy $\eta_j\ll 1$ ($j=x,y,z$) and the
trap frequency along the $x$ axis must satisfy $\nu_x\gg\kappa$.
Let us now consider a specific example from the ion-trapping
community: the trapped ion species ${}^9{\rm Be}^+$. 
Recent experiments with this particular ion (see, for example, 
\cite{Meekhof96,Monroe96}) have been performed with harmonic
oscillation frequencies along the principal axes of the trap 
$\nu_j/2\pi\simeq 11-30\;{\rm MHz}$, corresponding 
to Lamb-Dicke parameters $\eta_j\simeq 0.14-0.086$ 
(with respect to the $\,^2S_{1/2}\leftrightarrow\,^2P_{1/2}$
transition at wavelength $\lambda =313\;{\rm nm}$; the linewidth
for this transition is $\gamma /2\pi =19.4\;{\rm MHz}$).

A further practical consideration is the requirement that the 
spacing between the mirrors be large enough to accommodate the 
ion trap electrodes and external laser fields. A reasonable 
minimum separation might be $l=100\;\mu {\rm m}$, but we shall
make a somewhat more conservative choice of 
$l=250\;\mu {\rm m}$. With a cavity finesse ${\cal F}=300,000$
\cite{Mirrors},
one then obtains $\kappa /2\pi =1.0\;{\rm MHz}$.
Assuming, say, that $\eta_x=0.1$, then $\nu_x/2\pi =22\;{\rm MHz}$
and $\nu_x/\kappa =22$.
If the radius of curvature of the mirrors is taken to be
5 cm, then the cavity waist takes the value $w_0=15.8\;\mu {\rm m}$ 
(which yields $g_0/2\pi =14.9\;{\rm MHz}$) and 
$15\lambda^2{\cal F}/(2\pi^3w_0^2)=28\gg 1$. 

Finally, given these choices of parameters, a numerical estimate
for the rate $\Gamma =\Omega^2/\kappa$ at which the state transfer
occurs in the overdamped regime would be 
$\Gamma /2\pi\simeq 20-200\;{\rm kHz}$, and even 
larger with suitable laser pulse shapes in the underdamped regime.
This estimate establishes a timescale for the required stability 
of the driving laser fields (remember that the phases of the two
laser fields are assumed to be equal for the duration of the state
transfer process) and of the cavity and trap setups. 
Note that the timescales for motional decoherence and heating 
observed in recent trapped ion experiments (with ${}^9{\rm Be}^+$) 
are of the order of milliseconds and further improvement seems
possible \cite{Wineland98}.

So, it would seem that, with quite reasonable choices of
experimental parameters, a suitable operating regime for the
state transfer scheme is feasible. Note that favorable parameters
should also be achievable with trapped-ion species other than
${}^9{\rm Be}^+$; for example, ${}^{24}{\rm Mg}^+$ or
${}^{40}{\rm Ca}^+$. Alternatively, recent developments with
microscopic magnetic traps \cite{Weinstein95,Vuletic98}
suggest that suitably large trap frequencies and confinement 
in the Lamb-Dicke regime may also be possible with neutral 
atoms, such as Li and K. 

To conclude, in this work we have described a means of usefully
combining several emerging candidate technologies for the 
implementation of quantum communication and computing, i.e., 
trapped atoms, cavity QED, and propagating (nonclassical) light 
fields. 
The schemes outlined above allow, in principle, for the transfer
of quantum states and entanglement between light fields and
motional degrees of freedom of trapped atoms and for high-fidelity
transmission of quantum states between spatially distant sites.

\acknowledgments
We gratefully acknowledge the contributions of Prof. S. Braunstein,
whose insight was crucial in initiating the present work. 
ASP thanks A. Doherty, C. Hood, Q. Turchette, S. van Enk, L. You 
and P. Zoller for helpful discussions and the ITP in Santa Barbara
for its hospitality, where part of this work was carried out,
supported in part by the NSF under Grant No. PHY94-07194. 
ASP also thanks the Quantum Optics Group at Caltech for its 
hospitality and acknowledges support from the Marsden Fund of the 
Royal Society of New Zealand.
HJK is supported by the National Science Foundation, 
by DARPA via the QUIC Institute which is administered by ARO, 
and by the Office of Naval Research.

\appendix
\section{
Cascaded cavities correlation functions
}

Using Eq.~(\ref{Lc}), the equations of motion for the mean values
of the two cavity field amplitudes are straightforwardly derived as
\begin{eqnarray}
\langle\dot{\tilde{a}}_1\rangle &=& -\kappa \langle\tilde{a}_1\rangle ,
\\
\langle\dot{\tilde{a}}_2\rangle &=& -\kappa \langle\tilde{a}_2\rangle
- 2\kappa \langle\tilde{a}_1\rangle ,
\end{eqnarray}
for which the solutions are (for $t\geq 0$)
\begin{eqnarray}
\langle\tilde{a}_1(t)\rangle &=& e^{-\kappa t}
\langle\tilde{a}_1(0)\rangle , 
\\
\langle\tilde{a}_2(t)\rangle &=& e^{-\kappa t}
\langle\tilde{a}_2(0)\rangle
- 2\kappa t e^{-\kappa t} \langle\tilde{a}_1(0)\rangle .
\end{eqnarray}
These solutions are of the general form
\begin{equation}
\langle A_i(t)\rangle = \sum_j f_{ij}(t) \langle A_j(0)\rangle ,
\end{equation}
and the quantum regression theorem \cite{Gardiner91,Walls94} states
that two-time correlation functions follow as
\begin{equation}
\langle A_i(t)A_k(0)\rangle = \sum_j f_{ij}(t) \langle A_j(0)A_k(0)
\rangle .
\end{equation}
In steady state, the only nonzero equal-time correlations for our
system are 
\begin{equation}
\langle\tilde{a}_1\tilde{a}_1^\dagger\rangle_{\rm ss} =
\langle\tilde{a}_2\tilde{a}_2^\dagger\rangle_{\rm ss} = 1 ,
\end{equation}
which lead to the two-time correlation functions given in
(\ref{corr1}) and (\ref{corr2}).

\begin{figure}
\caption{
Schematic of experimental setup and excitation scheme for state
transfer between the motion of a trapped atom/ion and a quantized
cavity mode of the electromagnetic field.
}
\end{figure}

\begin{figure}
\caption{
Schematic of the cascaded system for transfer of statistics from 
a squeezed light field to the motional state of a trapped atom. 
Faraday isolators (F) facilitate a unidirectional coupling between 
the squeezed light source and the atom-cavity system.
}
\end{figure}

\begin{figure}
\caption{
Elements of the steady-state reduced density matrix of the motional 
mode of the trapped atom when the cavity mode is driven by quadrature 
squeezed light from a degenerate parametric oscillator. Parameters 
are given in the text.
}
\end{figure}

\begin{figure}
\caption{
Schematic of the cascaded system for transfer of a motional
state between separated trapped atoms/ions. The input to the first
cavity is ordinary vacuum and Faraday isolators (F) facilitate a
unidirectional coupling between the first and second cavities.
}
\end{figure}

\begin{figure}
\caption{
Fidelity of the transmission $F(t)$ for the state $|\psi\rangle$
of Eq.~(\ref{psi}) using the pulse shapes 
(\ref{Om_overdamped}) with $\Omega /\kappa =0.141$
($\Gamma /\kappa =0.02$), and taking $\nu /\kappa =20$ (solid), 
$10$ (dashed) and $5$ (dot-dashed).
}
\end{figure}

\begin{figure}
\caption{
Fidelity of the transmission $F(t)$ for the state $|\psi\rangle$
of Eq.~(\ref{psi}) for 
$\nu /\kappa =20$ with $\Omega /\kappa =0.4$ (solid), 
$0.5$ (dashed) and $0.7$ (dot-dashed), using the pulse shapes
(\ref{Om_overdamped}) (overdamped analysis).
The dotted line shows the result from the underdamped analysis
for $\Omega /\kappa =0.7$, 
using $\Omega_1(t)=\Omega$ for $t\geq 0$
and the form (\protect{\ref{eq:Om_underdamped}}) 
for $\Omega_1(t<0)$ [$\Omega_2(t)=\Omega_1(-t)$].
}
\end{figure}


\begin{thebibliography}{99}
\bibitem{Heinzen90b}
This follows the same principle as proposed in Ref.~\cite{Heinzen90} 
for cooling a source mode to its zero-point state by transferring its 
excitation to, in their case, a trapped ion, which is then cooled 
using sideband cooling.
\bibitem{LDlimit}
Note, however, that large squeezing implies population of large-$n$
number states and a broader spread of the atomic wave function,
which makes the Lamb-Dicke assumption of our scheme more restrictive.
\end{thebibliography}
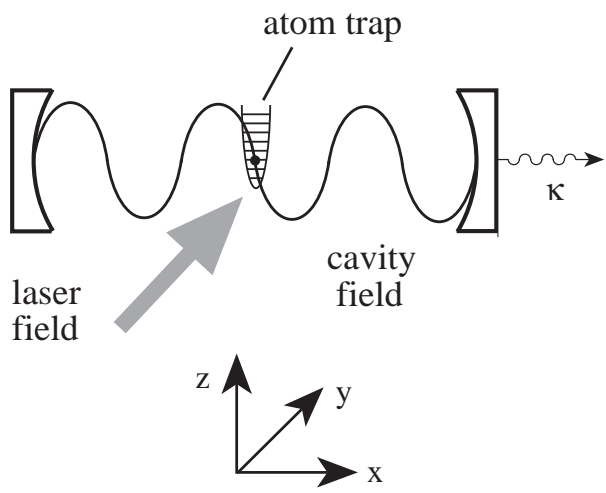
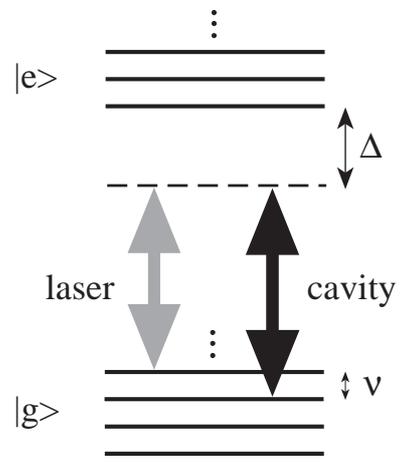
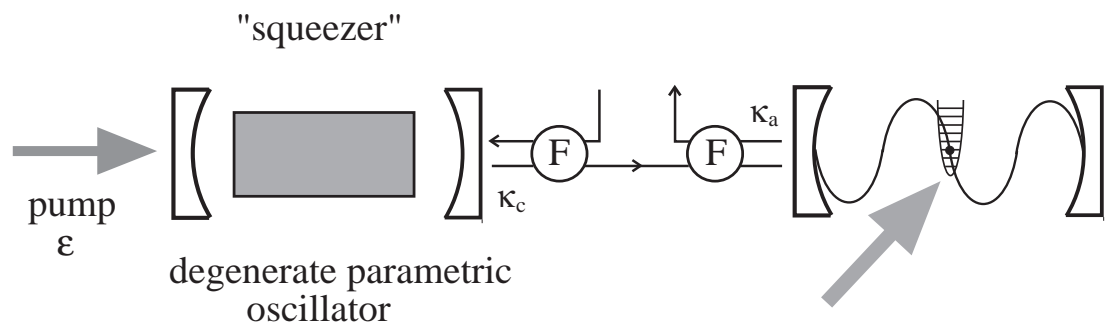
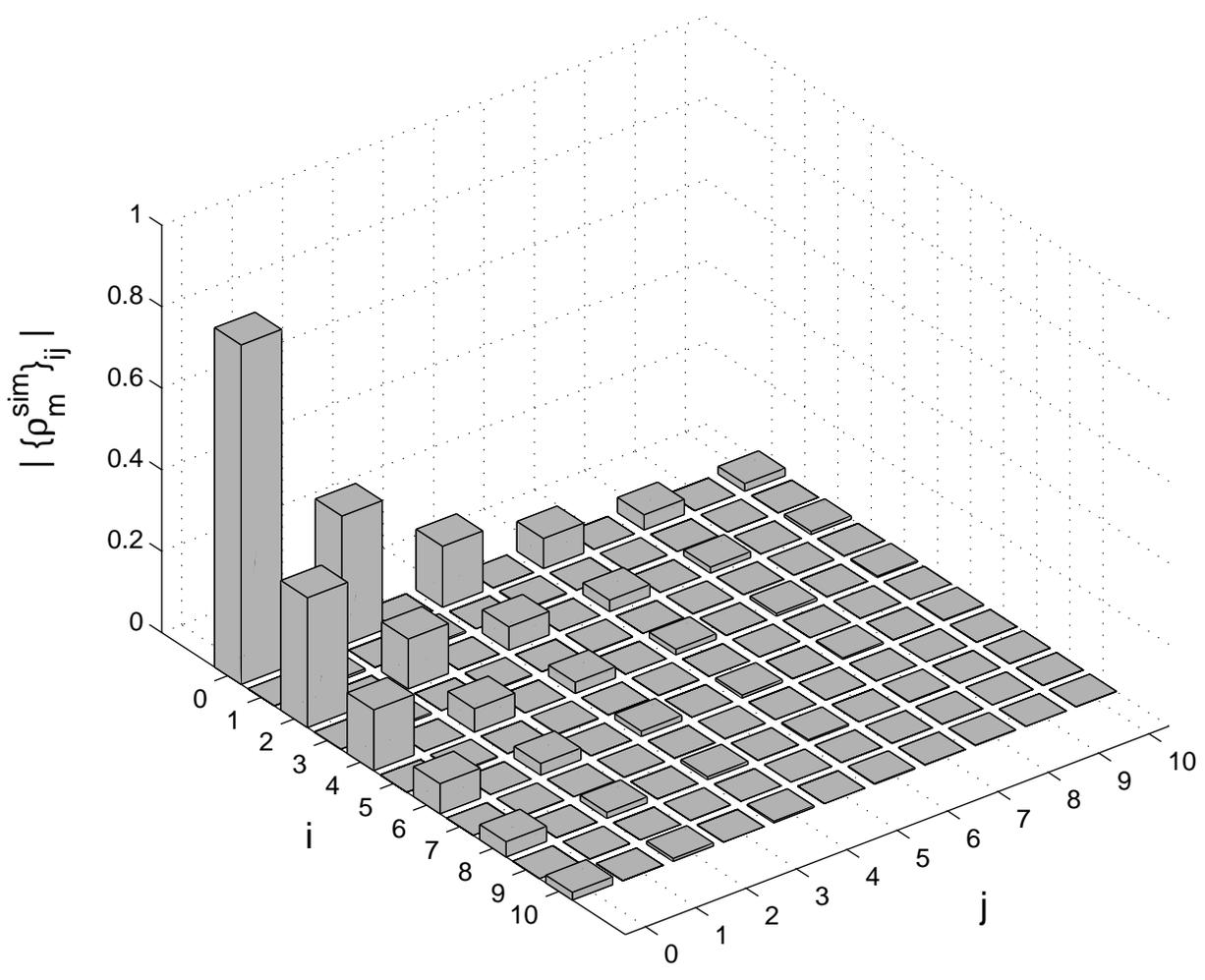
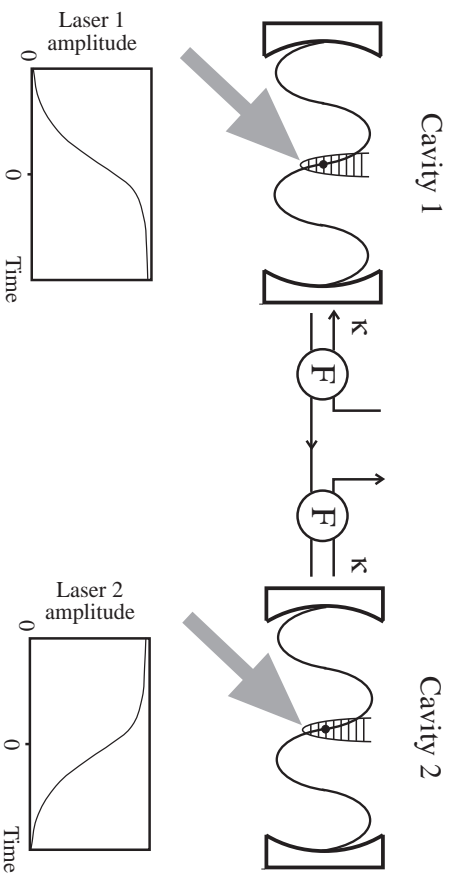
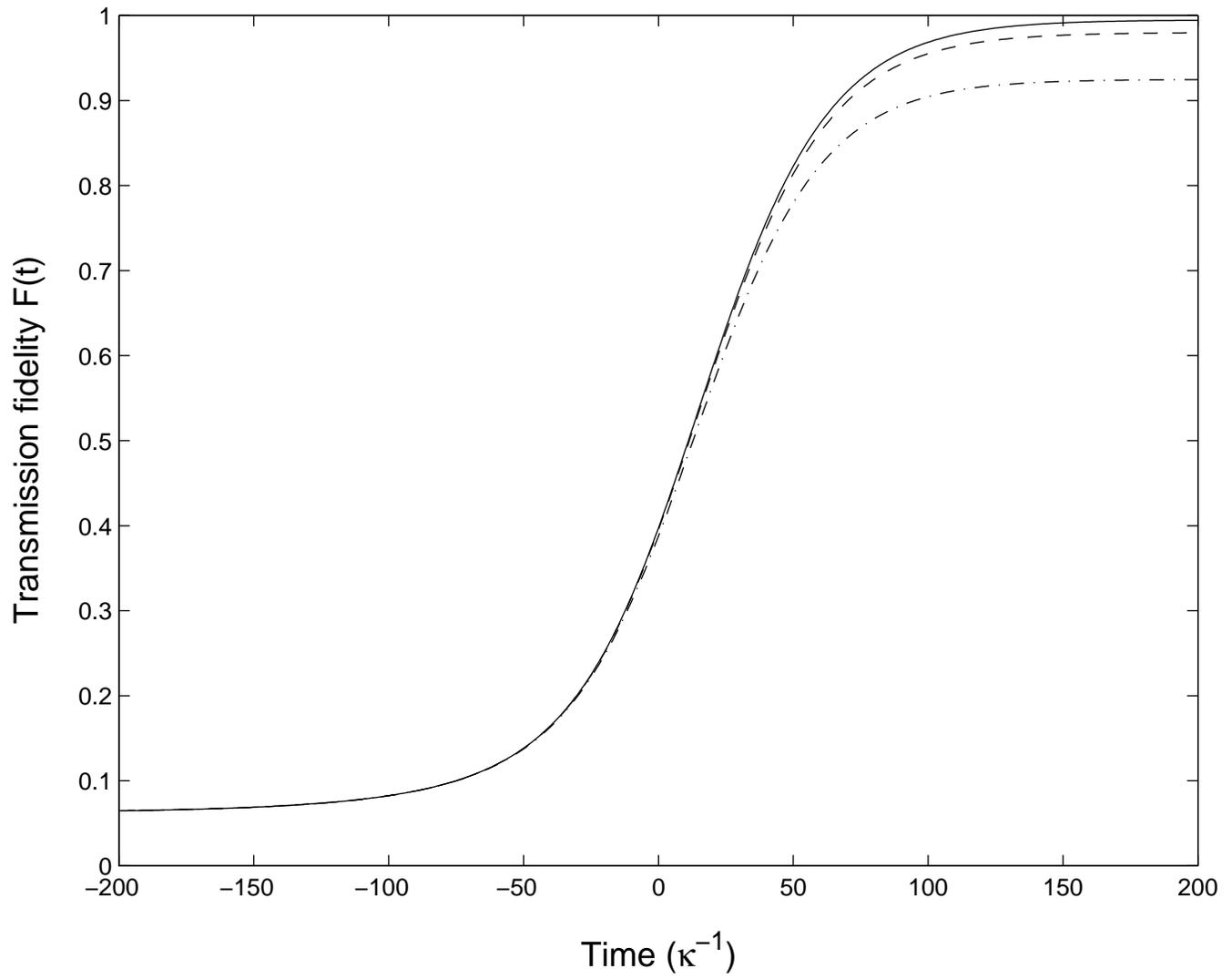
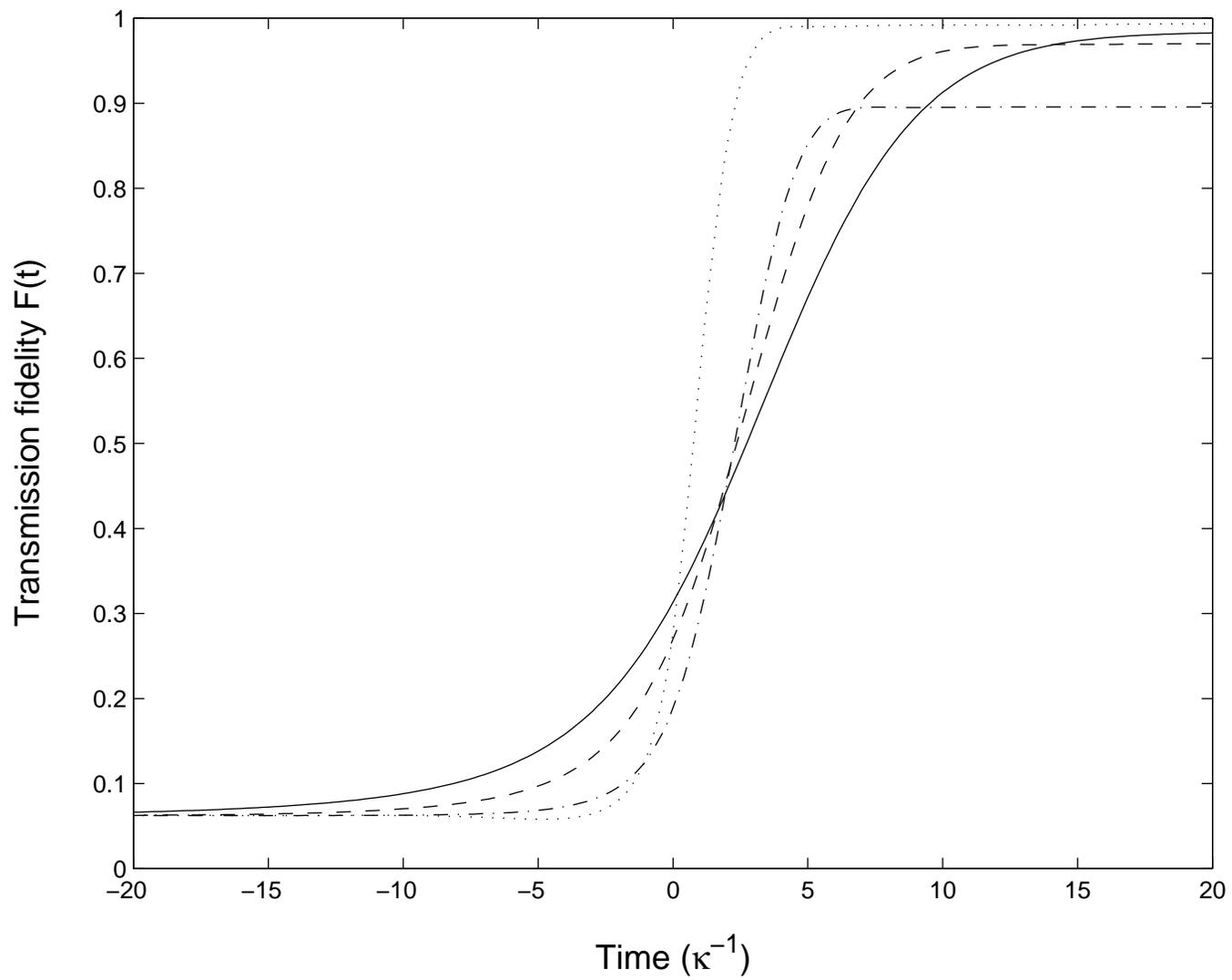
\end{document}